\documentclass[english,aps,preprint,showpacs]{revtex4}
\usepackage[T1]{fontenc}
\usepackage[latin1]{inputenc}
\usepackage{babel}
\usepackage{graphics}

\makeatletter

\providecommand{\LyX}{L\kern-.1667em\lower.25em\hbox{Y}\kern-.125emX\@}

\makeatother
\begin{document}

\title{Parallel versus sequential updating for Belief Propagation decoding}

\author{Haggai Kfir and Ido Kanter}

\affiliation{Minerva Center and Department of Physics, Bar-Ilan University, Ramat-Gan,
52900, Israel.}

\begin{abstract}
A sequential updating scheme (SUS) for the belief propagation algorithm
is proposed, and is compared with the parallel (regular) updating
scheme (PUS). Simulation results on various codes indicate that the
number of iterations of the belief algorithm for the SUS is about
one half of the required iterations for the PUS, where both decoding
algorithms have the same error correction properties. The complexity
per iteration for both schemes is similar, resulting in a lower total
complexity for the SUS. The explanation of this effect is related
to the inter-iteration information sharing, which is a property of
only the SUS, and which increases the \char`\"{}correction gain\char`\"{}
per iteration. 
\end{abstract}

\pacs{89.70.+C 89.20.Kk}

\maketitle

\section{Introduction }

Error correcting codes are essential part of modern communication,
enabling reliable transmission over noisy channels. Bounds over the
channels capacity were derived by Shannon in 1948 \cite{Shannon},
but more than four decades passed before codes that nearly saturate
the bound, such as Turbo code \cite{Turbo} and LDPC \cite{shokrolahi spilman irreg 65-85}
were presented. In recent years, an interesting bridge was established
between error correcting codes and statistical mechanics of disordered
systems \cite{Sourlas,sourlas - montanari,saad -kabashiba: SM of ECC}.

It is well known that as the noise level in a channel increases, the
decoding time (measured in algorithm iterations) also increases.\cite[section 3.2]{shokrollahi-richrdson-urbanke 3}.
Furthermore, as the noise \( f \) approaches the threshold level,
\( f_{c} \), the number of iterations diverges as a power-law, \( t\propto 1/(f_{c}-f) \)\cite{kanter saad construction}.
Hence, the acceleration of the decoding process, or the reduction
of the required number of iterations, is essential to ensure a smooth
information flow when operating near the channels capacity. 

In this paper we propose a variation of the well known Belief Propagation
(BP) algorithm \cite{mackay-neal-short-article4,Gallager ,Mackey good error 1},
which we label as \char`\"{}Sequential Updating Scheme\char`\"{}
(SUS). The complexity per iteration of the SUS is similar to the complexity
per iteration of the \emph{regular} Belief Propagation (BP) algorithm,
with Parallel Updating Scheme (PUS). However, simulations over a Binary
Symmetric Channel (BSC) indicate that for a given code, the SUS requires
about \textbf{\emph{}}\emph{one half} of the iterations in comparison
to the PUS, while the averaged bit error probability, \( p_{b} \),
is the same. 

This article is organized as follows: In section 2 the parallel and
the sequential updating schemes are defined. The distribution of the
decoding time obtained in simulations over BSC for both schemes are
presented in section 3. In section 4 the complexities of the sequential
and the parallel updating schemes are compared. A qualitative theoretical
argument supporting the acceleration of the decoding procedure in
the SUS is presented in section 5, and a detailed description of the
simulations is presented in section 6. A brief conclusion is presented
in section 7.

\section{The BP algorithm }

\subsection{Notation }

We focus on transmission over a BSC where each transmitted bit has
a chance \( f \) to flip during transmission, and a chance \( 1-f \)
of being transmitted correctly. Most of our results were obtained
using Mackey and Neal's algorithm (MN) , described in detail in reference
\cite{Mackey good error 1}. Briefly, the algorithm is defined as
follows: 

A word of size \( n \) is encoded into a codeword of size \( m \),
(\( Rate=n/m \)), using the following binary matrices: 

\begin{description}
\item [\( A: \)]a sparse matrix of dimensions \( (m\times n) \)
\item [\( B: \)]a sparse and invertible matrix of dimensions \( (m\times m) \). 
\end{description}
The encoding of a word \textbf{\( s \)} into a codeword \textbf{\( t \)}
is performed by: \begin{equation}
\label{MN_{e}ncode}
t=B^{-1}As\, (mod\: 2)
\end{equation}
 During the transmission, a noise \( n \) is added to the data, and
the received codeword \( r \) is:\begin{equation}
\label{r_{t}_{n}}
r=t+n\, (mod\, 2)
\end{equation}
 The decoding is performed by calculating \( z=B\cdot r \)\begin{equation}
\label{decoding}
z=B\cdot r=B\cdot (t+n)=B\cdot (B^{-1}\cdot A\cdot s+n)=A\cdot s+B\cdot n=[A,B][s,n],
\end{equation}

where {[} {]} represents appending matrices or concatenating vectors. 

Denoting \( H=[A,B] \) and \( x=[s,n] \), the decoding problem is
to find the most probable \textbf{\( x \)} satisfying: \( H\cdot x=z \)
(mod2) , where: 

\begin{description}
\item [\( H \)]is \( (m\times (n+m)) \) matrix 
\item [\( z \)]is the constraints (checks) vector of size \( m \). 
\item [\( x \)]is the unknown (variable) vector of size \( n+m \), termed
variable vector. 
\end{description}
This problem can be solved by a BP algorithm. Following \cite{Mackey good error 1,kanter - saad cascading 7},
we refer to the elements of \( x \) as nodes on a graph represented
by \( H \), the elements of \( z \) being values of checks along
the graph. The non-zero elements in a row \( i \) of \( H \) represent
the bits of \( x \) participating in the corresponding check \( z_{i} \).
The non-zero elements in column \( j \) represent the checks in which
the \( j \)th bit participates. 

We follow Kanter and Saad's (KS) construction for the matrices \( A,B \)
\cite{KS gaussian,kanter - saad cascading 7,kanter saad construction}.
This construction is characterized by very sparse matrices and a cyclic
form for \( B \), together with relatively high error correction
performance. 

For each non-zero element in \( H \), the algorithm calculates 4
values \cite{mackay-neal-short-article4}. The coefficient \( q^{1}_{ij} \)
(\( q^{0}_{ij} \)) stands for the probability that the bit \( x_{j} \)
is 1 (0), taking into account the information of all checks in which
it participates, except for the \( i \)th check. The coefficient
\( r_{ij}^{1} \) (\( r_{ij}^{0} \)) indicates the probability that
the bit \( x_{j} \) is 1 (0), taking into account the information
of all bits participating in the \( i \)th check, except for the
\( j \)th bit. 

The algorithm is initialized as follows. The coefficient \( Q_{j} \)
is set equal to our prior knowledge about that bit \( j \). In our
simulations we assume \( Q_{j}=0.5 \) if it is a source bit (\( j\leq N \)),
and \( Q_{j}=f \) if it is a noise bit (\( j>N \)). Then the \( q \)
values are set: \( q_{ij}^{1}=Q_{j} \) ; \( q_{ij}^{0}=1-q_{ij}^{1} \)
for all non-zero elements in the \( j \)th column.

\subsection{Parallel Updating Scheme (PUS)}

The PUS consists of alternating horizontal and vertical passes over
the \( H \) matrix. Each pair of horizontal and vertical passes is
defined as an iteration. In the horizontal pass, all the \( r_{ij} \)
coefficients are updated, row after row:\begin{equation}
\label{R0ij}
r^{0}_{ij}=\sum _{(all\, configurations\, with\, x_{j}=0,\, satisfing\, z_{i})}\prod _{j'\neq j}q_{ij'}^{x_{j'}}
\end{equation}
\begin{equation}
\label{R1ij}
r^{1}_{ij}=\sum _{_{}(all\, configurations\, with\, x_{j}=1,\, satisfing\, z_{i})}\prod _{j'\neq j}q_{ij'}^{x_{j'}}
\end{equation}
 where it is clear that the multiplication is performed only over
the non-zero elements of the matrix \( H \). 

A practical implementation of (\ref{R0ij}) and (\ref{R1ij}) is carried
out by computing the differences \( \delta q_{ij}\equiv q_{ij}^{0}-q_{ij}^{1} \),
and \( \delta r_{ij}\equiv r_{ij}^{0}-r_{ij}^{1} \) is then obtained
from the identity: \begin{equation}
\label{delta_{R}ij}
\delta r_{ij}=(-1)^{z_{i}}\prod _{j'\neq j}\delta q_{ij'}.
\end{equation}
 From the normalization condition \( r^{0}_{ij}+r^{1}_{ij}=1 \) one
can find: \begin{equation}
\label{extract_r01_from_delta_r}
r^{0}_{ij}=(1+\delta r_{ij})/2\, \, ;\, \, r^{1}_{ij}=(1-\delta r_{ij})/2
\end{equation}

(For a detailed description of this method, see \cite{Mackey good error 1}). 

In the vertical pass, all \( q^{1}_{ij},q^{0}_{ij} \) are computed,
column by column, using the updated values of \( r^{1}_{ij},r^{0}_{ij} \):\begin{equation}
\label{q0ij}
q_{ij}^{0}=\alpha _{ij}p_{j}^{0}\prod _{i'\neq i}r_{i'j}^{0}
\end{equation}
\begin{equation}
\label{q1ij}
q_{ij}^{1}=\alpha _{ij}p_{j}^{1}\prod _{i'\neq i}r_{i'j}^{1}
\end{equation}
 where \( \alpha _{ij} \) is a normalization factor, chosen to satisfy
\( q^{0}_{ij}+q^{1}_{ij}=1 \), and \( p_{j}^{0} \) , \( p_{j}^{1} \)
are the priors. Now the pseudo-posterior probability vector \textbf{\( Q \)}
can be computed by:\begin{equation}
\label{Qoj}
Q_{j}^{0}=\alpha _{j}p_{j}^{0}\prod _{i}r_{ij}^{0}
\end{equation}
\begin{equation}
\label{Q1j}
Q_{j}^{1}=\alpha _{j}p_{j}^{1}\prod _{i}r_{ij}^{1}
\end{equation}
 Again, \( \alpha _{j} \) is a normalization constant satisfying
\( Q^{1}_{j}+Q^{0}_{j}=1 \) , and \( i \) runs only over non-zero
elements of \( H \). Each iteration ends with a clipping of \textbf{\( Q \)}
to the variable vector \textbf{\( x \)} : if \( Q_{j}>0.5 \) then
\( x_{j}=1 \), else: \( x_{j}=0 \). At the end of each iteration
a convergence test, checking if \textbf{\( x \)} solves \textbf{\( Hx=z \)},
is performed. If some of the \( m \) equations are violated, the
algorithm turns to the next iteration until a pre-defined maximal
number of iterations is reached with no convergence (our halting criteria
are described in detail in section 6). Note that there is no inter-iteration
information exchange between the bits: all \( r_{ij} \) values are
updated using the previous iteration data.

\subsection{Sequential Updating Scheme (SUS)}

In the SUS, we perform the horizontal and vertical passes separately
for each bit in \( x \). A single sequential iteration for the bit
\( x_{j} \) consists of the following steps: 

\begin{enumerate}
\item For a given \( j \) all \( r_{ij} \) are updated. More precisely,
for all non-zero elements in column \( j \) of \( H \), use (\ref{delta_{R}ij},\ref{extract_r01_from_delta_r})
for updating \( r_{ij} \). Note that this is only a \emph{partial}
horizontal pass, since only \( r_{ij} \)'s belonging to a specific
column are updated. 
\item After all \( r_{ij} \) belonging to a column \( j \) are updated,
a vertical pass as defined in (\ref{q0ij},\ref{q1ij}) is performed
over column \( j \). Again, this is a \emph{partial} vertical pass,
referring only to one column. 
\item Steps 1-2 are repeated for the next column , until all columns in
\( H \) are updated. 
\item Finally, the pseudo posterior probability value \( Q_{j} \), is calculated
by (\ref{Qoj},\ref{Q1j}). 
\end{enumerate}
After all variable nodes are updated, the algorithm continues as for
the parallel scheme: clipping, checking the validity of the \( m \)
equations and proceeding to the next iteration.

\section{Results }

We performed simulations of decoding over a BSC using various rates,
block length and flip rates (\( f \)) and with the following constructions:
(a) KS \cite{kanter saad construction} construction; and (b) Irregular
LDPC codes, following the Luby - Mitzenmacher - Shokrollahi - Spielman
construction (LMSS), described in \cite{shokrolahi spilman irreg 65-85}.
Note that the LMSS code differs slightly from the MN code, but still
has a similar form of the BP scheme. We compare the distribution of
the convergence times of the PUS and SUS by decoding of the same codewords
(samples). 
\begin{figure}
{\centering \resizebox*{1\columnwidth}{10cm}{\rotatebox{270}{\includegraphics{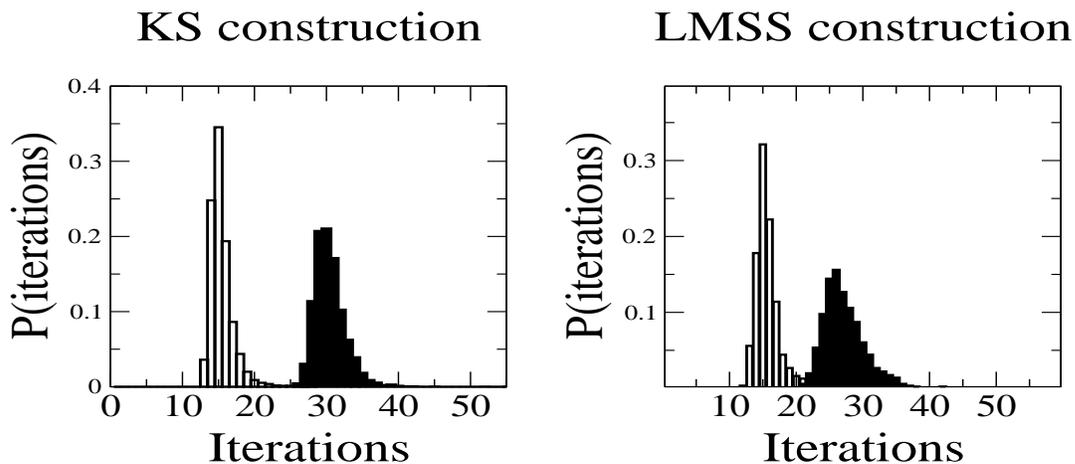}}} \par}

\caption{\label{figure 1}Distribution of the convergence times (measured
in iterations) for PUS and SUS (filled and empty bars, respectively),
for the KS and LMSS constructions. Rate is 1/2, \protect\protect\protect\( f=0.08\protect \protect \protect \)
and block size \protect\protect\( n=10,000\protect \protect \). }
\end{figure}

Figure \ref{figure 1} presents the distribution of the convergence
times (measured in algorithm iterations) for PUS (filled bars) and
SUS (empty bars) . The code rate is 1/2, \( f=0.08 \) (the channel
capacity is \( \approx 0.11 \) ) and the block length is \( 10,000 \).
The statistics were collected over at least \( 3,000 \) different
samples. The converging time for the SUS is about one half of the
converging time for the PUS. The average convergence time for the
PUS is 32.12 iterations for the KS construction; 28.52 for the LMSS
construction; while for the SUS the average convergence time is 16.7
and 16.32 iterations, respectively. It is worth mentioning that the
superior decoding time does not damage the error correcting property
of the code. In both constructions the observed bit error rate, \( p_{b} \),
after PUS or SUS, is nearly the same (see Table 1 for details). 
\begin{figure}
{\centering \resizebox*{1\columnwidth}{10cm}{\rotatebox{270}{\includegraphics{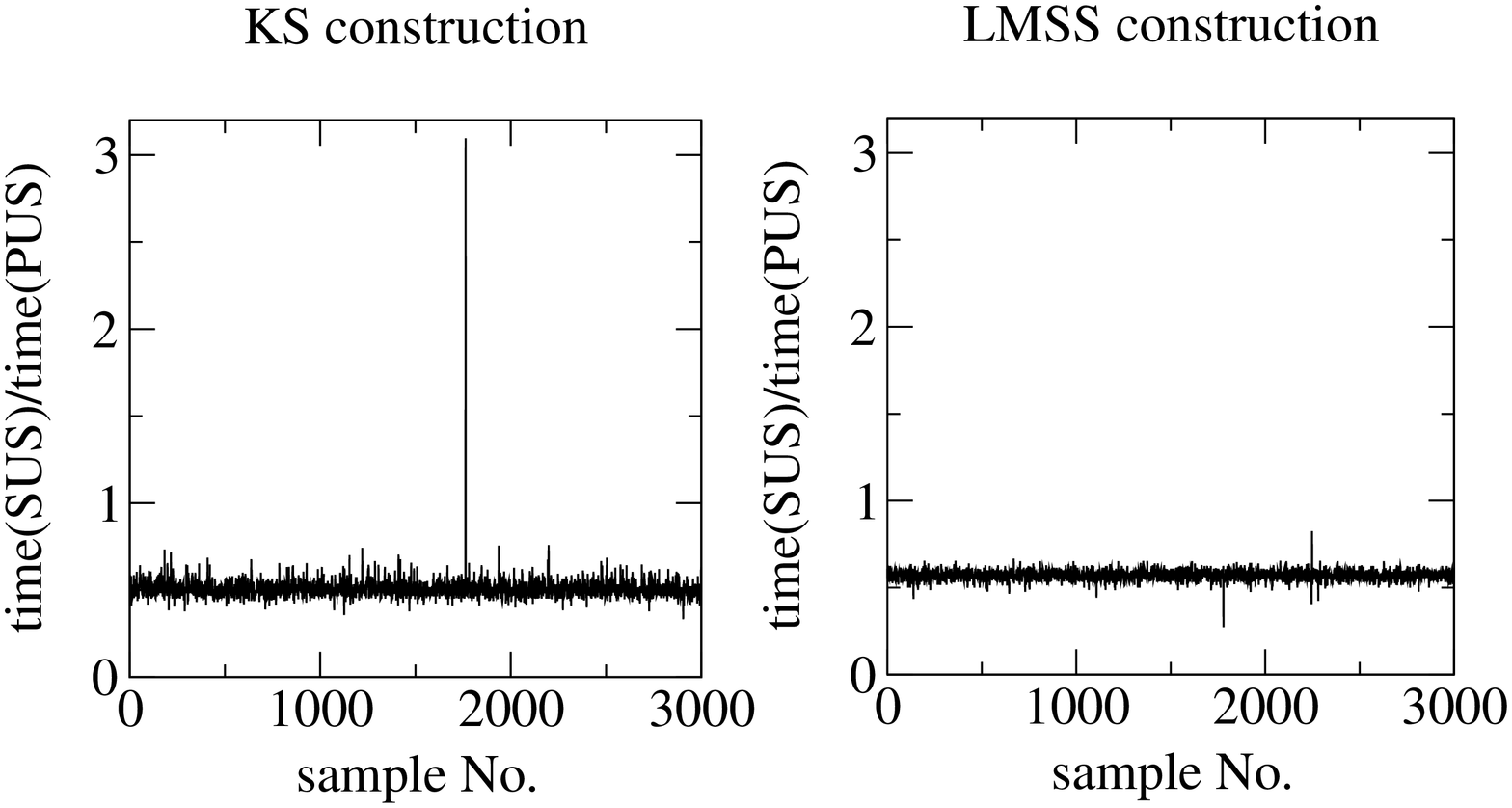}}} \par}

\caption{\label{figure 2}The ratio of convergence times, SUS time / PUS time,
per sample (time is measured in algorithm iterations). The rate is
almost a constant \protect\( (0.5)\protect \) independent of the
particular sample.}
\end{figure}

In Figure \ref{figure 2} the ratio between the converging times,
(SUS/PUS) per sample is plotted (time is measured in algorithm iterations).
For the vast majority of the samples this is very close to the average
rate. This indicates that the double number of iterations for the
PUS in comparison to the SUS is the typical result. 

Table 1 presents similar measurements for other rates and noise levels.
Results indicate the following general rule. Independent of the construction,
the noise level and the rate, the convergence time of the PUS is around
double the number of iterations required to achieve convergence in
the SUS. (As our statistics was collected over \( \sim 3000 \) samples
of block size \( 10^{4} \), we do not report the exact value for
\( p_{b}\leq 10^{-5} \) ). 

\begin{tabular}{|c|c|c|c|c|p{2cm}|}
\hline 
construction&
 \( \begin{array}{c}
<t_{PUS}>\\
(iterations)
\end{array} \)&
 \( \begin{array}{c}
<t_{SUS}>\\
(iterations)
\end{array} \)&
 \( \left\langle \frac{t_{SUS}}{t_{PUS}}\right\rangle  \)&
 \( \frac{\left\langle t_{SUS}\right\rangle }{\left\langle t_{PUS}\right\rangle } \)&
 bit error rate (PUS, SUS)\\
\hline
\multicolumn{1}{|l|}{KS,R=1/2, \( f \)=0.09}&
 58.3&
 28.4&
 0.488&
 0.504&
\multicolumn{1}{l|}{\( 5.4\cdot 10^{-5} \), \( 5.5\cdot 10^{-5} \)}\\
\hline
\multicolumn{1}{|l|}{KS, R=1/2, \( f \)=0.08}&
 32.12&
 16.7&
 0.519 &
 0.5076&
\multicolumn{1}{l|}{\( p_{b}\leq 10^{-5},p_{b}\leq 10^{-5} \)}\\
\hline
\multicolumn{1}{|l|}{KS, R=1/2, \( f \)=0.07}&
 23.0&
 11.54&
 0.505&
 0.501&
\multicolumn{1}{l|}{\( p_{b}\leq 10^{-5},p^{b}\leq 10^{-5} \)}\\
\hline
\multicolumn{1}{|l|}{KS, R=1/3, \( f \)=0.159}&
 65.58&
 32.54&
 0.496&
 0.5236&
\multicolumn{1}{l|}{\( 3.4\cdot 10^{-4},2.08\cdot 10^{-4} \)}\\
\hline
\multicolumn{1}{|l|}{KS,R=1/3, \( f \)=0.15}&
 40.53&
 21.28&
 0.525&
 0.525&
\multicolumn{1}{l|}{\( p_{b}\leq 10^{-5},p_{b}\leq 10^{-5} \)}\\
\hline
\multicolumn{1}{|l|}{KS, R=1/5, \( f \)=0.23}&
 103.9&
 54.66&
 0.526&
 0.5005&
\multicolumn{1}{l|}{\( 1.9\cdot 10^{-3},1.9\cdot 10^{-3} \)}\\
\hline
\multicolumn{1}{|l|}{KS, R=1/5, \( f \)=0.21}&
 36.98&
 19.46&
 0.527&
 0.526&
\multicolumn{1}{l|}{\( p_{b}\leq 10^{-5},p_{b}\leq 10^{-5} \)}\\
\hline
\multicolumn{1}{|l|}{LMSS, R=1/2, \( f \)=0.08}&
 28.52&
 16.32&
 0.572&
 0.573&
\multicolumn{1}{l|}{\( p_{b}\leq 10^{-5},p_{b}\leq 10^{-5} \)}\\
\hline
\multicolumn{1}{|l|}{LMSS,R=1/2, \( f \)=0.07}&
 17.90&
 10.94&
 0.611&
 0.612&
\multicolumn{1}{l|}{\( p_{b}\leq 10^{-5},p_{b}\leq 10^{-5} \)}\\
\hline
\end{tabular}

\section{Complexity per iteration }

In this section we show that the complexity per single iteration is
almost the same for both methods. Hence, the gain in iterations represents
the gain in decoding complexity. 

For both schemes, the calculation of the pseudo posterior probabilities,
the clipping and the convergence tests are identical, so they can
be excluded from our discussion. Furthermore, the vertical passes
in the two schemes are identical, hence the only remaining source
for a possible difference in the complexity for the two schemes is
the horizontal pass. For simplicity, we assume a regular matrix \textbf{\( H \)}
of dimensions \( m \) rows by \( n \) columns, which has \( k \)
nonzero elements per row and \( c \) nonzero elements per column.
(One can easily extend the discussion to include irregular matrices,
but the conclusions are the same). 

In the PUS, each horizontal pass consists of \( k \) subtraction
operations to find \( \delta q_{ij} \)'s and \( (k-1) \) multiplications
to find \( \delta r_{ij} \)'s (using (\ref{delta_{R}ij})) for each
bit. The calculation of \( r^{1}_{ij} \) \& \( r^{0}_{ij} \) from
\( \delta r_{ij} \) requires two additions and two multiplications
(\ref{extract_r01_from_delta_r}). The total number of operations
per iteration for the PUS is given by 

\begin{description}
\item [additions:]\begin{equation}
\label{par aditions}
m(k+2k)=3mk
\end{equation}

\item [multiplications:]\begin{equation}
\label{par mltiplications}
m(k(k-1)+2k)=mk(k+1)
\end{equation}

\end{description}
In the SUS, horizontal passes are done separately for each bit, summing
to \( n\cdot c \) passes in total. Each horizontal pass consists
of \( k-1 \) subtractions to find \( \delta q_{ij'} \) for all participants
in the check, except for the current bit, and \( k-1 \) multiplications
are required to calculate \( \delta r_{ij} \). The calculation of
\( r^{1}_{ij} \) \& \( r^{0}_{ij} \) from \( \delta r_{ij} \) in
this scheme requires two additions and two multiplications per bit.
Hence the total complexity is given by 

\begin{description}
\item [additions:]\begin{equation}
\label{seq addition}
nc(k-1+2)=nc(k+1)
\end{equation}

\item [multiplications:]\begin{equation}
\label{seq multiplication}
nc(k-1+2)=nc(k+1)
\end{equation}

\end{description}
Recalling that \( mk=nc \), we have: 

\begin{itemize}
\item (\ref{par mltiplications}) equals (\ref{seq multiplication}), so
the same number of multiplications is performed in both cases. 
\item (\ref{seq addition}) becomes \( m(k^{2}+k) \), which for \( k>2 \)
is larger than (\ref{par aditions}). However, for small \( k \)
( \( k\leq  \)5 for KS, and \( \left\langle k\right\rangle  \) is
of the same order for LMSS), the increment in the total fraction of
additions is small. Furthermore, one must remember that \emph{the
complexity is dominated by the number of multiplications}. 
\end{itemize}
Note that the abovementioned comparison was made under a straightforward
implementation of both algorithms. In advanced algorithms the following
improvements can be adopted in order to reduce the complexity of the
schemes. 

\begin{enumerate}
\item Some savings can be made for the horizontal passes in the PUS, for
instance, computing \( \prod \delta q_{ij} \) for the entire row,
and dividing by each \( \delta q_{ij} \) element, or recomputing
\( \delta q_{ij} \) only for updated bits in the SUS. 
\item PUS can be implemented simultaneously over all checks (variables)
using several processors. The implementation of SUS in parallel over
a finite fraction of the checks (variables) is possible, but may require
a special design. 
\item SUS has some advantage in memory requirement, since only a column
vector of the currently updated \( r_{ij} \) is required, whereas
for the PUS the whole \( r_{ij} \) matrix must be retained simultaneously. 
\end{enumerate}

\section{Qualitative theoretical explanation}

The key difference between the two algorithms is the inter iteration
information exchange, which is a property of the SUS only. Let us
denote by \( r^{t}_{ij},q^{t}_{ij}\equiv  \) the values computed
in iteration \( t \). In the PUS all \( r_{ij}^{t} \) values are
determined by the \( q_{ij}^{t-1} \) values (values of the previous
iteration), and the \( q_{ij}^{t} \)'s are determined by these \( r_{ij}^{t} \)'s
. In the SUS, after a bit is updated, the following bits that share
a check with it are already exposed to the updated information. For
instance, assume \( x_{j} \) and \( x_{k} \) share a check \( i \),
i.e. \( H_{ij}=H_{ik}=1 \), and assume \( j<k \). In iteration \( t \),
\( r_{ij}^{t} \) is updated using \( q_{ik}^{t-1} \); however, proceeding
to the \( k \)th column, \( r_{ik}^{t} \) is updated using \( q_{ij}^{t} \),
the most recent available information. 

In other words, in the SUS, the first bits to be updated utilize the
previous iteration data \( (q_{ij}^{t-1}) \). A group of bits use
mixed data from previous and current iterations and, finally, some
of the bits are entirely updated by information from the current step
\( (q_{ij}^{t}) \). 

The gain in the number of iterations for the SUS can be qualitatively
well understood by the following argument: Since the decoding procedure
terminates successfully (with some small \( p_{b} \)) and the number
of correct bits increases monotonically, it is evident that on average,
the current knowledge is superior to the knowledge of previous iteration.%
\footnote{Our simulations indicate that the number of correct bits versus time
(iterations) is a \emph{concave function} (except for the last few
iterations). This result leads to a faster convergence for the SUS
(many-small-steps converging strategy) in comparison to the PUS.
}

An important question is raised regarding which part of the SUS is
accelerated in comparison to the PUS. The acceleration of the SUS
may be a result of one of the following regimes: (a) a faster asymptotic
convergence; (b) a faster arrangement in the initial stage of the
decoding from random initial conditions; or (c) a uniform acceleration
over all the stages of the decoding. 

\begin{figure}
{\centering \resizebox*{1\columnwidth}{10cm}{\rotatebox{270}{\includegraphics{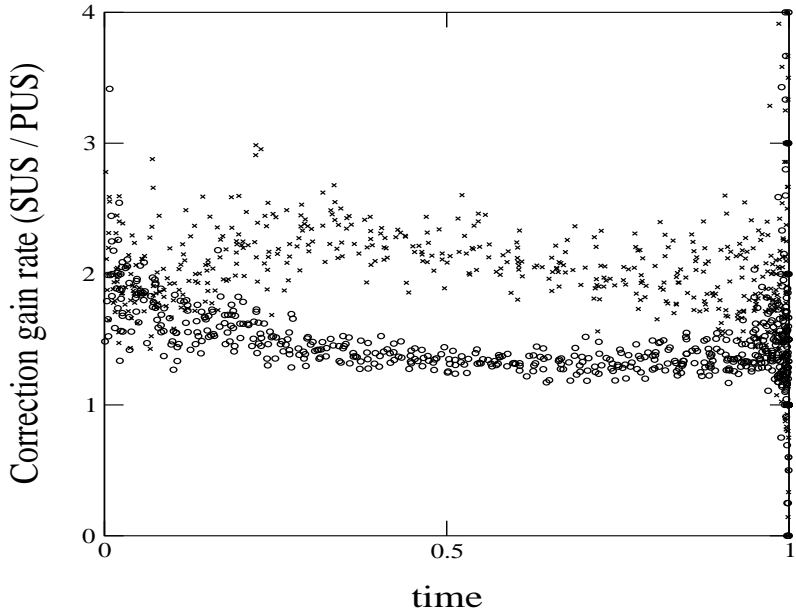}}} \par}

\caption{\label{graph 3}The rate between the correction gain of SUS and PUS
versus the time (iterations). The data was collected over each iteration
of a KS construction, rate \protect\( 1/3,f=0.155\protect \) and
block size \protect\( 10,000\protect \). The symbol {}``\protect\protect\protect\( \times \protect \protect \protect \)''
represents a forward updating order, and {}``\protect\protect\protect\( \circ \protect \protect \protect \)''
represents a reverse updating order. The correction gain rate is almost
time independent, and is higher for forward updating due to the properties
of the KS construction.}
\end{figure}

In order to answer this question we perform the following advanced
simulations. We run the PUS and record the number of correct bits
in each iteration. The correction gain of each iteration is defined
as the increment in the fraction of correct bits. At each step of
the PUS we prepare another replica of the system with the same initial
conditions, the same \( q_{ij} \) and \( r_{ij} \), and run one
iteration of the SUS. The correction gain of the SUS is then compared
to that of the PUS. In Figure \ref{graph 3} we plot the rate between
the sequential and parallel correction gains as a function of time
(marked \( \times  \)). This rate is nearly \( 2 \), with relatively
small fluctuations along the decoding process. In other words, on
the average the SUS corrects twice the number of bits in comparison
to the PUS, independent of the state of the decoder. (The simulation
was performed on the KS construction with rate \( 1/3 \), \( f=0.155 \),
block size \( 10000 \), \( 20 \) different samples, and all convergence
times were normalized to a 0-1 scale.) \par{}

The observation that the correction gain uniformly distributed over
all the stages of the decoding raises the question of whether there
is a superior updating order of the bits producing a correction gain
greater than \( 2 \). For one iteration of the KS construction, it
may be better to update the bits from left to right than in the reverse
order. For rate \( 1/3 \), for instance, the right-most part of the
matrix (about 25\% of the columns) contain only one non-zero element
per column, and this element is also the last element in its check.
These bits are entirely updated by current iteration data, resulting
in an increased correction gain. At the left-most end, on the other
hand, there are 7 non-zero elements per column (for rate 1/3 construction),
so that only a small fraction of them are the \char`\"{}last bit\char`\"{}
for all the checks in which they participate. Most of the bits are
updated by mixed information from the current and previous iteration,
resulting in a smaller correction gain. In Figure \ref{graph 3} the
rate between the correction gain for SUS and PUS for reversed (right
to left) updating order is marked {}``\( \circ  \)''. This rate
is evidently less than for the left to right updating. Preliminary
simulations indicate that by carefully selecting the updating order
one can save 10\%-30\% of the iterations relative to a plain left
to right sequential updating.

\section{Simulations}

In this sections the technical details of our simulations are described.
We generate the \( H \) matrix at random, distributing the non-zero
elements as evenly as possible without violating the constraint of
the number of elements per row/column. No special attempt was made
to select a \char`\"{}good performing\char`\"{} matrix. For the
KS structure, we generate the \textbf{\( x \)} vector as follows:
The source bits were set to \( 1 \) or \( 0 \) with probability
0.5. The noise bits were set to \( 0 \), and then exactly a fraction
\( f \) of the bits were selected randomly and flipped (\( f \)
is the flip probability). The check vector \textbf{\( z \)} was computed
by \( z=Hx \), and the algorithm solved \( Hx'=z \). We found \( p_{b} \)
by comparing \textbf{\( x \)} \& \textbf{\( x' \)}, for the source
region only. The source length selected was \( n=10,000 \) (resulting
in \textbf{\( x \)} of length \( 40,000 \), and \textbf{\( z \)}
of length \( 30,000 \) for rate \( 1/3 \)). 

For the LMSS structure, following \cite{shokrolahi spilman irreg 65-85}
we always decoded the all-zero codeword, generating the noise vector
\textbf{\( n \)} in the same way as described above. The check vector
\textbf{\( z \)} was computed, \( z=Hn \), and the algorithm solved
\( Hn'=z \). We found \( p_{b} \) by comparing \textbf{\( n \)}
\& \textbf{\( n' \)} (in the LMSS version the \char`\"{}decoding\char`\"{}
ends when the noise vector is found and the transmitted vector, \textbf{\( t \)},
is related to the received vector, \textbf{\( r \)} by \( t=r+n \)(mod
2). Finding the source message from \( t \) is not defined as part
of the decoding problem). We used a noise vector of length \( 20,000 \),
corresponding to a check vector of size \( 10,000 \) (rate 1/2). 

In both cases the flip rate, \( f \), was selected as being close
enough to the critical rate for this block length such that the decoding
is characterized by relatively long convergence times. However, the
flip rate \( f \) was chosen not too close to the threshold in order
to avoid a large fraction of non-converging samples. After the check
vector \textbf{\( z \)} was constructed, it was decoded both in parallel
and sequential schemes, and the number of iterations was monitored.
We defined 3 halting criteria for the iterative process: 

\begin{enumerate}
\item The outcome \textbf{\( x' \)} fully solves \textbf{\( Hx'=z \)}. 
\item The algorithm reached a stationary state, namely, \textbf{\( x' \)}
did not change over the last 10 iterations. 
\item A predefined number of iteration was exceeded (\char`\"{}non-convergence\char`\"{}).
This number was selected so as to be far larger than the average converging
time (500 iterations in our case). 
\end{enumerate}
The vast majority of samples converged successfully. More precisely,
less than 0.2\% samples failed to converge or reach a non-solving
stationary state.

\section{Conclusions }

We demonstrated that the SUS outperforms the PUS in the convergence
time aspect by a factor of about 2. Since the complexity per iteration
of the two schemes is nearly the same, the gain in iterations is similar
to the gain in the decoding complexity. The time gain is probably
related to the inter-iteration information exchange, which is a property
of the SUS. This explanation is also consistent with the observation
that the gain is uniformly distributed over all the decoding stages.
The question of whether the number of iterations can be reduced by
a factor greater than \( 2 \) by updating the bits in a special order
is currently under investigation. 

We acknowledge fruitful discussions with D. Ben-Eli and I. Sutskover.

\end{document}